\documentclass{ptephy_v1}

\usepackage{amsmath,amssymb}
\usepackage{graphicx}
\usepackage{float}

\def\({\left(}
\def\){\right)}
\def\d{\partial}
\newcommand\bs[1]{\boldsymbol{\mathit{#1}}}

\renewcommand{\iint}{\int\!\!\!\!\int}
\renewcommand{\iiint}{\int\!\!\!\!\int\!\!\!\!\int}

\begin{document}

\title{Distributions of Consecutive Level Spacings of
Gaussian Unitary Ensemble and Their Ratio: ab initio Derivation}

\author{\name{\fname{Shinsuke} \surname{M. Nishigaki}}{\ast}} 

\address{\affil{~}{
Graduate School of Natural Science and Technology,
Shimane University, Matsue 690-8504, Japan}
\email{nishigaki@riko.shimane-u.ac.jp}}

\begin{abstract}
In recent studies of 
many-body localization in nonintegrable quantum systems,
the distribution of the ratio of two consecutive energy level spacings,
$r_n=(E_{n+1}-E_n)/(E_{n}-E_{n-1})$ or $\tilde{r}_n=\min(r_n,r_n^{-1})$,
has been used as a measure to quantify the chaoticity,
alternative to the more conventional distribution of the level spacings,
$s_n=\bar{\rho}(E_n)(E_{n+1}-E_n)$,
as the former makes unnecessary the unfolding required for the latter.
Based on our previous work on the Tracy-Widom approach to the J\'{a}nossy densities,
we present analytic expressions for the joint probability distribution
of two consecutive eigenvalue spacings
and the distribution of their ratio for the Gaussian unitary ensemble (GUE) of 
random Hermitian $N\times N$ matrices at $N\to \infty$,
in terms of a system of differential equations.
As a showcase of the efficacy of our results for characterizing an approach to quantum chaoticity,
we contrast them to arguably the most ideal of all quantum-chaotic spectra: the zeroes
of the Riemann $\zeta$ function on the critical line at increasing heights.
\end{abstract}

\subjectindex{A10, A13, A32, B83, B86}

\maketitle

\noindent
{\it 1. Introduction.}
Many-body localization
that prohibits thermal equilibration of the wave functions has been a core agenda of research
in the field of quantum many-body systems, including
disordered and interacting fermions on a chain \cite{Oganesyan:2007};
spin chains with transverse field, periodical kicks, or disorder
\cite{Atas:2013,Kim:2013,DAlessio:2014,Luitz:2015};
and the Sachdev-Ye-Kitaev model and its deformed, coupled, or sparsed variant
\cite{You:2017,Garcia:2018,Garcia:2019,Garcia:2021},
to name but a few.
In all of the above studies, the ``gap ratio distribution," i.e.
the distribution $P_{\mathrm{r}}(r)$ of the ratio of consecutive energy level spacings
$r_n=(E_{n+1}-E_n)/(E_{n}-E_{n-1})$ or $\tilde{r}_n=\min(r_n,r_n^{-1})$,
initiated by Ref.\,\cite{Oganesyan:2007},
has been utilized as a criterion of (non)ergodicity.
This trend is obviously due to its computational advantage
that makes unnecessary the unfolding by the smoothed density of states $\bar{\rho}(E)$,
over the conventional distribution $P(s)$ of level spacings
$s_n=\bar{\rho}(E_n)(E_{n+1}-E_n)$.
More recently, the use of the gap ratio distribution
for characterizing spectral transitions extended its reach beyond
energy level statistics of many-body or chaotic Hamiltonians,
namely towards quantum entanglement spectra of reduced density matrices
in neural network states \cite{Deng:2017}
and in quantum circuits \cite{Zhang:2020,Jian:2022}.

The random matrix theory of the gap ratio distribution was introduced
in Ref.\,\cite{Atas:2013} and has since been quoted in practically every work in these fields
including Refs.\,\cite{Kim:2013,DAlessio:2014,Luitz:2015,You:2017,Garcia:2018,%
Garcia:2019,Garcia:2021,Deng:2017,Zhang:2020,Jian:2022}.
There, two types of approximate expressions for $P_{\mathrm{r}}(r)$ were presented:
one from the Wigner-like surmise
(substituting the large-$N$ limit of $N\times N$ matrices by $3\times 3$)
for all three Dyson symmetry classes,
$P_{\mathrm{r}}(r)\approx C_\beta (r+r^2)^\beta/(1+r+r^2)^{1+3\beta/2}\ (\beta=1,2,4)$,
and another from the quadrature discretization of the
resolvent operator $\bs{K}_I(\mathbb{I}-\bs{K}_I)^{-1}$
where $\bs{K}_I$ denotes the integral operator of 
convolution with the sine kernel over an interval $I$ for the unitary class $\beta=2$.
Although the latter approximation is known to converge to the exact value quickly
as the quadrature order is increased \cite{Bornemann:2010a}, 
the analytical expression for $P_{\mathrm{r}}(r)$ is still missing.
Moreover, the author cannot help feeling frustrated to find that every article
that quotes Ref.\,\cite{Atas:2013} almost always refers only to the Wigner-surmised form,
and calls that crude, uncontrolled approximation the ``outcome from the random matrix theory".
In view of these, this article aims to determine analytically the joint distribution of
consecutive eigenvalue spacings $P_{\mathrm{c}}(a,b)$ and the distribution of their ratio
$P_{\mathrm{r}}(r)$ for the unitary class,
based on our recent work \cite{Nishigaki:2021} which provided a generic prescription for
determining the J\'{a}nossy density for any integrable kernel
as a solution to the Tracy-Widom (TW) system of 
partial differential equations (PDEs) \cite{Tracy:1994c}.

This article is composed of the following parts: 
In Sect.\,2 we list several facts on the J\'{a}nossy density $J_1(0;I)$ for the sine kernel,
i.e.~the conditional probability that an interval $I$ in the spectral bulk of the GUE
contains no eigenvalue except for the one at a designated locus.
In Sect.\,3 we follow the prescription of TW
and show that the J\'{a}nossy density and 
associated distributions of consecutive eigenvalue spacings and their ratio are
analytically determined as a solution to a system of ordinary differential equations (ODEs).
As a showcase of the efficacy of our results for characterizing
an approach to quantum chaoticity without using $\bar{\rho}(E)$,
in Sect.\,4 we contrast them to the distribution of zeroes of the Riemann $\zeta$
function on the critical line at increasing heights.
A program for generating the J\'{a}nossy density is
attached as Supplementary material.

{\it 2. J\'{a}nossy density for the sine kernel.}
The sine kernel 
\begin{align}
K(x,y)&=\frac{\sin(x-y)}{\pi(x-y)}
=\frac{\sin x \cos y-\cos x \sin y}{\pi(x-y)}
\label{Ksine}
\end{align}
governs the determinantal point process of unfolded\footnote{
We adopt a normalization such that the mean eigenvalue spacing is $\pi$.}
eigenvalues or eigenphases
$\{x_i\}$ of random Hermitian or unitary matrices of infinite rank $N\to \infty$,
in the spectral bulk \cite{Mehta,Forrester}.
From the very defining property of
the determinantal point process that the $p$-point correlation function 
${R}_p(x_1,\ldots,x_p)=\mathbb{E}[ \prod_{i=1}^p\delta(x-x_i)]$
is given by $\det[K(x_i, x_j)]_{i,j=1}^p$,
it follows that the {\em conditional} $p$-point correlation function
\begin{align}
\tilde{R}_{p|1}(x_1,\ldots,x_p|t)=
\frac{\mathbb{E}[\delta(x-t)\prod_{i=1}^p\delta(x-x_i)]}{\mathbb{E}[\delta(x-t)]},
\end{align}
in which one of the eigenvalues is preconditioned at $x=t$,
also takes a determinantal form
$\det[\tilde{K}(x_i, x_j)]_{i,j=1}^p$
governed by another kernel \cite{Nishigaki:2021},
\begin{align}
\tilde{K}(x,y)=K(x,y)-K(x,t)K(t,t)^{-1}K(t,y).
\label{SL2R}
\end{align}
In the case of the sine kernel (\ref{Ksine}),
its translational invariance allows for setting $t=0$ without loss of generality,
so that \cite{Nagao:1993},
\begin{align}
\tilde{K}(x,y)&=
\frac{1}{\pi}\(\frac{\sin(x-y)}{x-y}-\frac{\sin x}{x}\frac{\sin y}{y}\)
=\frac{\varphi(x)\psi(y)-\psi(x)\varphi(y)}{x-y}~,
\nonumber
\\
\varphi(x)&=\frac{1}{\sqrt{\pi}}\sin x ,~~
\psi(x) =\frac{1}{\sqrt{\pi}}\(\cos x-\frac{\sin x}{x}\).
\label{Ktil}
\end{align}
The transformation of kernels (\ref{SL2R}) from $K$ to $\tilde{K}$ 
is associated with a meromorphic $\mathrm{SL}(2,\mathbb{R})$ gauge transformation
\cite{Nishigaki:2021},
\begin{align}
\left[
\begin{array}{c}
\varphi(x)\\
\psi(x)
\end{array}
\right]
=
\left[
\begin{array}{cc}
1&0\\
-x^{-1}&1\\
\end{array}
\right]
\left[
\begin{array}{c}
\pi^{-1/2}\sin x\\
\pi^{-1/2}\cos x
\end{array}
\right],
\end{align}
on the two-component functions that comprise respective kernels
in the right-hand sides of Eqs.\,(\ref{Ksine}) and (\ref{Ktil}).
Accordingly, as stated in Theorem in Ref.\,\cite{Nishigaki:2021},
the gauge-transformed section $[\varphi(x), \psi(x)]^T$ 
inherits from the original section $[\pi^{-1/2}\sin(x), \pi^{-1/2}\cos(x)]^T$
the covariant-constancy condition for a meromorphic
$\mathfrak{sl}(2,\mathbb{R})$ connection, 
i.e. a pair of linear differential equations,
\begin{align}
m(x)\frac{d}{dx}
\left[
\!
\begin{array}{c}
\varphi(x)\\
\psi(x)
\end{array}
\!
\right]
=
\left[
\!
\begin{array}{rr}
A(x) \!&\!\! B(x)\\
-C(x) \!&\!\! -A(x)
\end{array}
\!
\right]
\left[
\!
\begin{array}{c}
\varphi(x)\\
\psi(x)
\end{array}
\!
\right]
\,\mbox{with polynomials}~
m(x), A(x), B(x), C(x),
\label{integrable}
\end{align}
which guarantees applicability of the TW method.
In the present case of spherical Bessel functions (\ref{Ktil}), 
the polynomials comprising the connection (Lax operator)
$\frac{1}{m}\left[ {~A~~B \atop -C\,-A}\right]$
are:
\begin{align}
A(x)=1,~~B(x)=C(x)=m(x)=x.
\label{ABCm}
\end{align}
Subsequently we shall use their nonzero coefficients,
\begin{align}
\alpha_0=\beta_1=\gamma_1=\mu_1=1.
\label{coeffs}
\end{align}

We take an interval $I=[a_1, a_2]$ with $a_1<0<a_2$
so that the ordered triple  $(a_1, 0, a_2)$ will serve as three consecutive eigenvalues,
and denote by $\tilde{\bs{K}}_I$ the integration operator
acting on the Hilbert space of square-integrable functions $L^2(I)$ with
convolution kernel $\tilde{K}(x,y)$,
\begin{align}
(\tilde{\bs{K}}_I f)(x):=\int_I dy\,\tilde{K}(x,y)f(y).
\end{align}
Then by Gaudin and Mehta's theorem \cite{Mehta},
the J\'{a}nossy density $J_1(0;[a_1,a_2])$ \cite{Borodin:2003}, i.e.~the
conditional probability that the interval $I=[a_1, a_2]$
contains no eigenvalue except for the one preconditioned at $x=0$, is
expressed as the Fredholm deterninant of $\tilde{\bs{K}}_I$:
\begin{align}
J_1(0;[a_1,a_2])&=
\mathrm{Det}\,(\mathbb{I}-\tilde{\bs{K}}_I)=
\exp\Bigl(-\sum_{n\geq 1}\frac1n\mathrm{Tr}\,\tilde{\bs{K}}_I^n\Bigr)
\nonumber
\\
&=
\exp\(
-\int_I dx\,\tilde{K}(x,x)
-\frac12 \iint_I dx dy\,\tilde{K}(x,y)\tilde{K}(y,x)
-\cdots\).
\label{FredholmDet}
\end{align}
Note that the J\'{a}nossy density for a symmetric interval $I=[-t,t]$
was previously expressed in terms of a Painlev\'{e} V transcendent,
i.e. a special solution to an
ODE in $t$ \cite{Forrester:1996}.
Our task in this article is to extend their result to a generic interval
and express $J_1(0;[a_1,a_2])$ 
and an associated distribution $P_{\mathrm{r}}({r})$
in terms of a system of PDEs in $a_1$ and $a_2$.

{\it 3. TW system.}
TW \cite{Tracy:1994c} established a systematic method of
computing the Fredholm determinant
of an integrable integral kernel whose component functions satisfy
the condition in Eq.\,(\ref{integrable}).
The quantities that appear in the TW system
[$j,k=1$ or $2$],
\begin{align}
R_{jk}&=
\tilde{K}(a_j, a_k)+
\int_I dx \, \tilde{K}(a_j, x)\tilde{K}(x, a_k)+
\iint_I dx dy\, \tilde{K}(a_j, x)\tilde{K}(x, y)\tilde{K}(y, a_k)+\cdots
\nonumber\\
&=((\mathbb{I}-\tilde{\bs{K}}_I)^{-1}\tilde{{K}})(a_j, a_k)=R_{kj},
\nonumber\\
q_j&=
\varphi(a_j)+\int_I dx\,\tilde{K}(a_j, x) \varphi(x)
+\iint_I dx dy\,\tilde{K}(a_j, x)\tilde{K}(x, y)  \varphi(y)+
\cdots
\nonumber\\
&=((\mathbb{I}-\tilde{\bs{K}}_I)^{-1}  \varphi)(a_j),
\nonumber\\  
p_j&=((\mathbb{I}-\tilde{\bs{K}}_I)^{-1}  \psi)(a_j),
\nonumber\\  
u&=
\int_I dx\,\varphi(x)^2
+\iint_I dx dy\,\varphi(x)\tilde{K}(x, y)  \varphi(y)
+\iiint_I dx dy dz\,\varphi(x)\tilde{K}(x, y)\tilde{K}(y, z)  \varphi(z)+
\cdots
\nonumber\\
&=
\int_I dx\, \varphi(x) ((\mathbb{I}-\tilde{\bs{K}}_I)^{-1} \varphi)(x),
\nonumber\\
v&=\int_I dx\, \psi(x) ((\mathbb{I}-\tilde{\bs{K}}_I)^{-1} \varphi)(x),~~
w=\int_I dx\, \psi(x) ((\mathbb{I}-\tilde{\bs{K}}_I)^{-1}\psi)(x),
\label{vars}
\end{align}
are all treated as functions of the left and the right endpoints $(a_1, a_2)$
of the interval $I$.
Expanding the definitions in Eqs.\,(\ref{FredholmDet}) and (\ref{vars})
in $a_1$ and $a_2\ll1$, 
the boundary conditions for these quantities read:
\begin{align}
\ln  J_1(0;[a_1,a_2])&=\frac{a_1^3-a_2^3}{9\pi}
-\frac{2 (a_1^5-a_2^5)}{225\pi}+\cdots,
\nonumber\\
q_j&=
\frac{a_j}{\sqrt{\pi}}-\frac{a_j^3}{6\sqrt{\pi}}
-\frac{(a_1^3-a_2^3)a_j}{9\pi^{3/2}}+\frac{a_j^5}{120\sqrt{\pi}}+\cdots,
\nonumber\\
p_j&=
-\frac{a_j^2}{3\sqrt{\pi}}+\frac{a_j^4}{30\sqrt{\pi}}
+\frac{(a_1^4-a_2^4) a_j}{36\pi^{3/2}}+\cdots,
\nonumber\\
u&=-\frac{a_1^3-a_2^3}{3\pi}+\frac{a_1^5-a_2^5}{15\pi}+\cdots,~~
v=\frac{a_1^4-a_2^4}{12\pi}+\cdots,~~
w=-\frac{a_1^5-a_2^5}{45\pi}+\cdots,
\label{bc}
\end{align}
up to terms of $O(a_1, a_2)^6$.
Substituting the coefficients (\ref{coeffs}) into 
the TW system of PDEs (Eqs.\,(2.25), (2.26), (2.31), (2.32), (2.12)-(2.18), (1.7a) of 
Ref.\cite{Tracy:1994c}), it takes the following form
[below the pair of indices $(j, k)$ assumes either $(1, 2)$ or $(2, 1)$]:
\begin{align}
R_{jk} &=\frac{q_j p_k-p_j q_k}{a_j-a_k}~,
\nonumber\\
a_j \frac{\d q_j}{\d a_j}&=U q_j+(V+a_j)p_j-(-1)^k a_k R_{jk}q_k~,
\nonumber\\
a_j \frac{\d p_j}{\d a_j}&=-U p_j+(V-a_j)q_j-(-1)^k a_k R_{jk}p_k~,
\nonumber\\
\frac{\d q_j}{\d a_k}&=(-1)^k R_{jk}q_k~,~~
\frac{\d p_j}{\d a_k }=(-1)^k R_{jk}p_k~,
\nonumber\\
\frac{\d u}{\d a_j}&=(-1)^j q_j^2~,~~
\frac{\d v}{\d a_j} =(-1)^j q_jp_j~,~~
\frac{\d w}{\d a_j} =(-1)^j p_j^2~,
\label{TWsystem}
\end{align}
where $U=1+u-w$ and $V=2v$.
The ``stiffness" of the second and the third equations of Eqs.\,(\ref{TWsystem})
at $a_j=0$ is only superficial, because  $q_j$, $p_j$, and $R_{jk}$
are of $O(a_j)$ or higher orders in the limit $a_j\to 0$.
We remark that the above PDEs reduce to the ODEs
(Eqs.\,(14), (15), (17)) of Ref.\,\cite{Forrester:1996} and $V=0$
in the symmetric case, $|a_1|=a_2=t$.
The Fredholm determinant (\ref{FredholmDet}) is expressed by $R_{jj}$,
which is composed of $q_j$ and $p_j$ (Eqs.\,(1.3), (1.7b) of Ref.\,\cite{Tracy:1994c}),
\begin{align}
(-1)^{j-1} \frac{\d}{\d a_j}\ln J_1(0;[a_1,a_2])
=R_{jj}
=p_j \frac{\d q_j}{\d a_j}-q_j \frac{\d p_j}{\d a_j}~~~(j=1,2).
\label{Rjj}
\end{align}

For numerical evaluation of $J_1(0;[a,b])$,
it is practically convenient to start from the initial condition (\ref{bc}) at
$(\epsilon a, \epsilon b)$ with a sufficiently small $\epsilon>0$ and
to integrate the TW system (Eqs.\,(\ref{TWsystem}, \ref{Rjj}))
in the radial direction $\(a_1(s), a_2(s))=(sa, sb\)$, $\epsilon\leq s\leq 1$.
The resultant system of ODEs in $s$, combined from the PDEs through
$s\frac{d}{ds}=a_1(s) \frac{\d}{\d a_1}+a_2(s)\frac{\d}{\d a_2}$,
reads:
\begin{align}
&s\frac{d}{ds}
\left[
\begin{array}{l}
q_j\\
p_j
\end{array}
\right]
=
\left[
\begin{array}{cc}
U& V+a_j\\
V-a_j &-U
\end{array}
\right]
\left[
\begin{array}{l}
q_j\\
p_j
\end{array}
\right]~,
\nonumber\\
&s\frac{dU}{ds}=-a_1(q_1^2-p_1^2)+a_2(q_2^2-p_2^2)~,~
s\frac{dV}{ds}=-2a_1 q_1 p_1+2a_2 q_2 p_2~,
\nonumber\\
&s\frac{d\ln J_1}{ds}=a_1(q_1^2+p_1^2)-a_2(q_2^2+p_2^2)-(q_1 p_2-p_1 q_2)^2
+2U(q_1 p_1-q_2 p_2)
\nonumber\\
&~~~~~~~~~~~~-V(q_1^2-p_1^2-q_2^2+p_2^2)~.
\label{ode}
\end{align}
Note that the form of the first line of Eq.\,(\ref{ode}) is directly inherited from
Eqs.\,(\ref{integrable}) and (\ref{ABCm}).
It is quite plausible that the ODEs (\ref{ode}) can be expressible as a Hamiltonian system,
and $J_1(0; [a_1,a_2])$ be regarded as a $\tau$-function of an integrable hierarchy
\cite{Forrester:2001}.
\begin{figure}[t] 
\begin{center}
\includegraphics[bb=0 0 567 373,width=76mm]{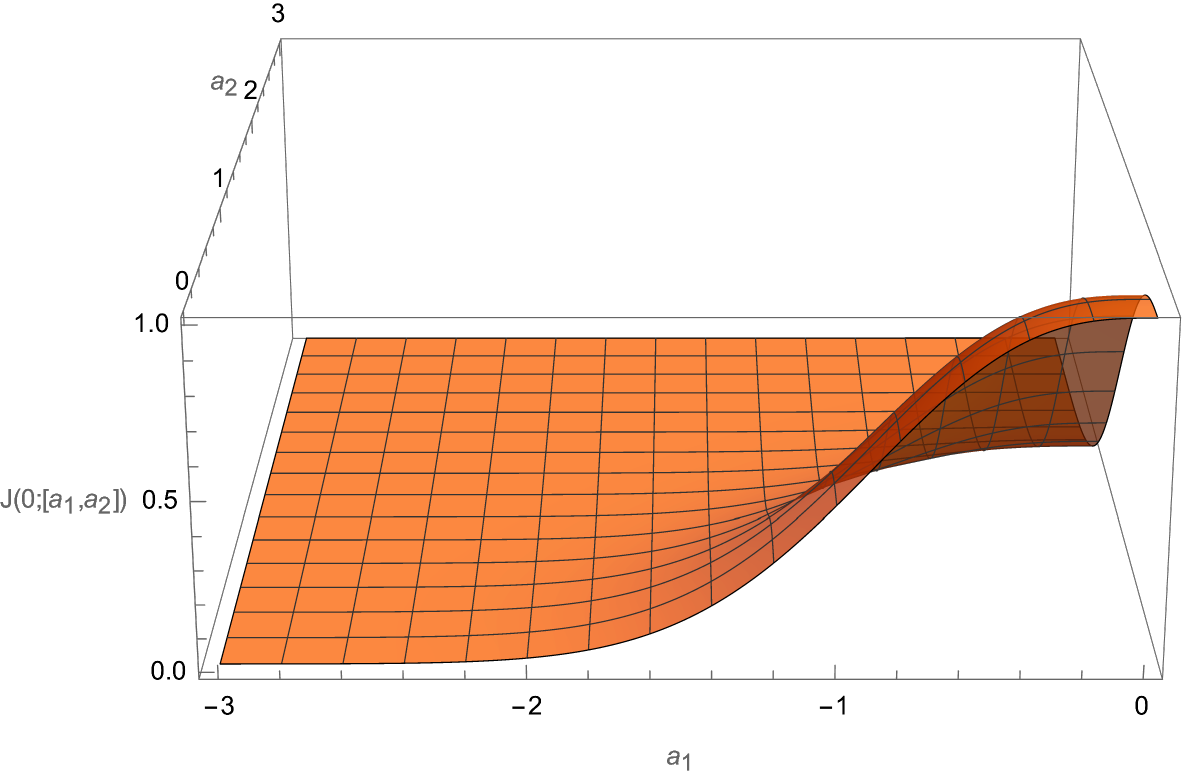}
\includegraphics[bb=0 0 573 378,width=76mm]{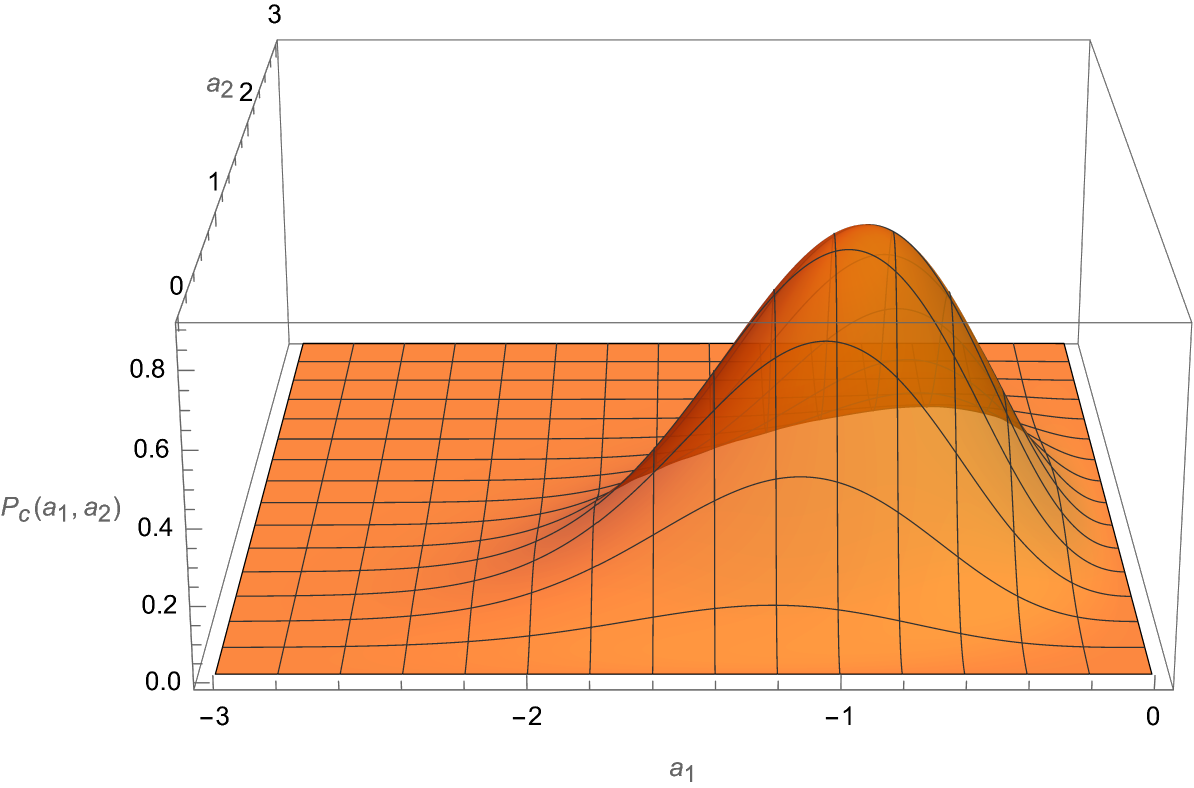}
\caption{
The J\'{a}nossy density $J_1(0;[a_1,a_2])$ for the sine kernel (left), and
the joint distribution $P_{\mathrm{c}}(a_1,a_2)$ of two consecutive eigenvalue spacings (right).
For visual clarity, each  coordinate $a_j$ is rescaled from the text by $1/\pi$,
so that the mean eigenvalue spacing is unity.
}
\end{center}
\end{figure}
However, these are not immediately evident to the author, 
and will be discussed in a separate publication.
As a cross-check of our formulation,
we confirmed that the values of $J_1(0;[a_1,a_2])$ obtained by the above prescription
(Fig.\,1, left) are identical,
within the accuracy of numerical evaluation,
to those from the Nystr\"{o}m-type quadrature approximation
of the Fredholm determinant \cite{Bornemann:2010a}
\begin{align}
\mathrm{Det}\,(\mathbb{I}-\tilde{\bs{K}}_I)
\simeq \det\left[ \delta_{ij} -\tilde{K}(x_i ,x_j)\sqrt{w_i w_j} \right]_{i,j=1}^m~,
\label{Nystrom}
\end{align}
where $\{x_i, w_i\}_{i=1}^m$ is the $m$-th order quadrature of the interval $I$ such that 
$\sum_{i=1}^m w_i f(x_i)\stackrel{m\to\infty}{\longrightarrow}\int_I dx\,f(x)$,
with a sufficiently large $m$.
Specifically, relative deviations of $J_1(0;[a_1,a_2])$
computed by the TW system (\ref{ode}) starting from the initial value $\epsilon= 10^{-10}$
using Mathematica's {\tt NDSolve} package
with {\tt WorkingPrecision}\,$\to$\,5\,{\tt MachinePrecision},
and that evaluated by the Nystr\"{o}m-type approximation (\ref{Nystrom}) with
the Gauss-Legendre quadrature of order $m=200$,
do not exceed $10^{-27}$ for the whole range of variables $|a_j| \leq 10$,
and $<10^{-19}$ for $|a_j| \leq 20$.
Interested readers are invited to verify this statement by running the
Notebook {\tt Janossy\_TW\_N.nb} included in Supplementary materials.

Various probability density functions follow from the J\'{a}nossy density:
The level spacing distribution $P(s)$ \cite{Mehta},
the distribution for the nearest neighbor level spacing
$P_{\mathrm{nn}}(t)$ \cite{Forrester:1996},
and the joint distribution for the two consecutive level spacings $P_{\mathrm{c}}(a_1, a_2)$ 
\cite{Atas:2013}
are given by
\begin{align}
P(s)=-\frac{dJ_1(0;[0,s])}{ds},\ \ 
P_{\mathrm{nn}}(t)&=-\frac{dJ_1(0;[-t,t])}{dt},\ \ 
P_{\mathrm{c}}(a_1,a_2)=
-\frac{\d^2 J_1(0;[a_1,a_2]) }{\d a_1 \d a_2},
\label{Pca1a2}
\end{align}
respectively (Fig.\,1, right).
Finally, the distribution $P_{\mathrm{r}}(r)$ for the ratio $r=|a_1|/a_2$
of the two consecutive level spacings is given by \cite{Atas:2013},
\begin{align}
P_{\mathrm{r}}(r)&=
\int_0^\infty da_2 \int_{-\infty}^0 da_1 P_{\mathrm{c}}(a_1,a_2)\,
\delta(r-|a_1|/a_2)=
\int_0^\infty da\, a\, P_{\mathrm{c}}(-r a,a)~.
\label{Prr}
\end{align}
If we switch the variable from $r$ to 
$\tilde{r}:=\min(|a_1|, a_2)/\max(|a_1|, a_2)=\min(r, r^{-1})\in [0,1]$,
its distribution is twice  the above $P_{\mathrm{r}}(r)$ (Fig.\,2), 
yielding the expectation values for its moments,
\begin{align}
\mathbb{E}[\tilde{r}^k]
&=2\int_0^\infty da_2 \int_{-a_2}^{0} da_1 P_{\mathrm{c}}(a_1,a_2) \(|a_1|/a_2\)^k
\nonumber\\
&=
0.5997504209(1),\,
0.4132049292(1),\,
0.3100223500(1),\,
0.2460560527(1)
\nonumber\\
&~~~~(k=1,2,3,4).
\label{Er}
\end{align}

\begin{figure}[t] 
\begin{center}
\includegraphics[bb=0 0 450 287,width=75mm]{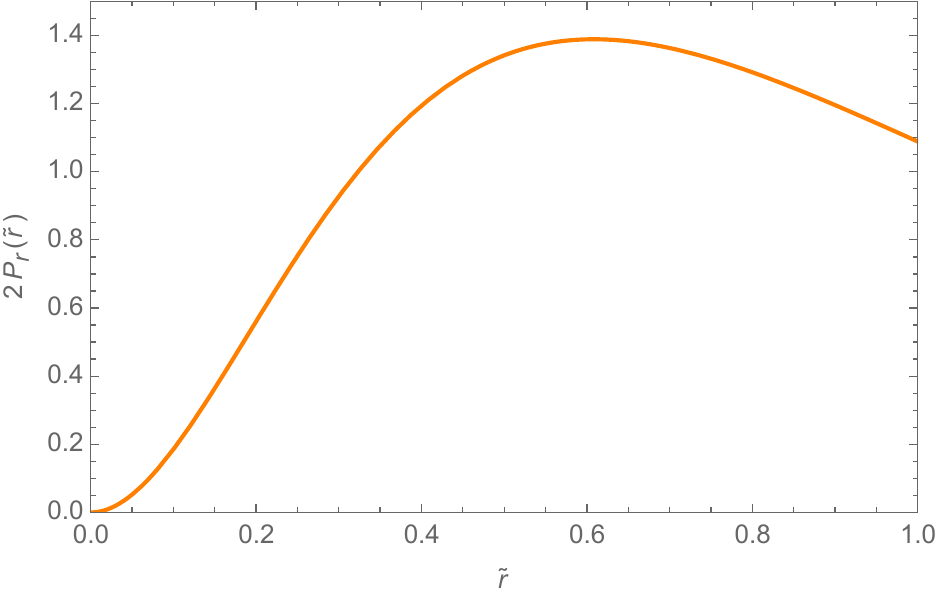}
\caption{
The distribution of the $\tilde{r}$-ratio,
$\tilde{r}=\min(|a_1|, a_2)/\max(|a_1|, a_2)$.}
\end{center}
\end{figure}

{\it 4. Zeroes of the Riemann $\zeta$ function.}
As a showcase of application of our analytic results to the judgement of quantum chaoticity,
we compare them to (arguably) the most ideal of all quantum-chaotic spectra:
the sequence of zeroes of the Riemann $\zeta$ function on the critical line,
$\{\frac12+ i\gamma_n\}$. 
Their imaginary parts are supposed to be the eigenvalues of
the hypothetical self-adjoint Hilbert-P\'{o}lya operator \cite{Wiki:Hilbert-Polya}.
Extensive computational and analytic number theory studies
since Montgomery's pair correlation conjecture \cite{Montgomery:1972}
have fruited in conviction that
such an operator, if interpreted as a Hamiltonian,
should be ergodic and possess no
antiunitary convolutive symmetry, i.e. belong to the unitary universality class
of random matrices \cite{Odlyzko:1987,Rudnick:1996}.
After unfolding by the Riemann-von Mangoldt formula for the asymptotic density of zeroes,
$\bar{\rho}(\gamma)=\frac{1}{2\pi}\log \frac{\gamma}{2\pi}$,
the histogram $P_{\mathrm{c}}(\delta_-,\delta_+)$ of two unfolded consecutive spacings 
$\delta_\pm=\bar{\rho}(\gamma_n)(\gamma_{n\pm 1}-\gamma_{n})$
of $10^8$ zeroes ending at $n=103800788359$
 (the largest zero available at the L-functions and modular forms database \cite{LMFDB})
perfectly agree with the GUE result (\ref{Pca1a2}),
as visualized in Fig.\,3 (left).
Moreover, 
{\em had we not known the classical formula for $\bar{\rho}(\gamma)$ a priori,}
the perfect match to the GUE could still be deduced from
the distributions of the ratios of consecutive spacings of zeroes
$r_n=(\gamma_{n+1}-\gamma_{n})/(\gamma_{n}-\gamma_{n-1})$
and $\tilde{r}_n=\min(r_n, r_n^{-1})$ (Fig.\,3, right).
\begin{figure}[t] 
\begin{center}
\includegraphics[bb=0 0 780 492,width=76mm]{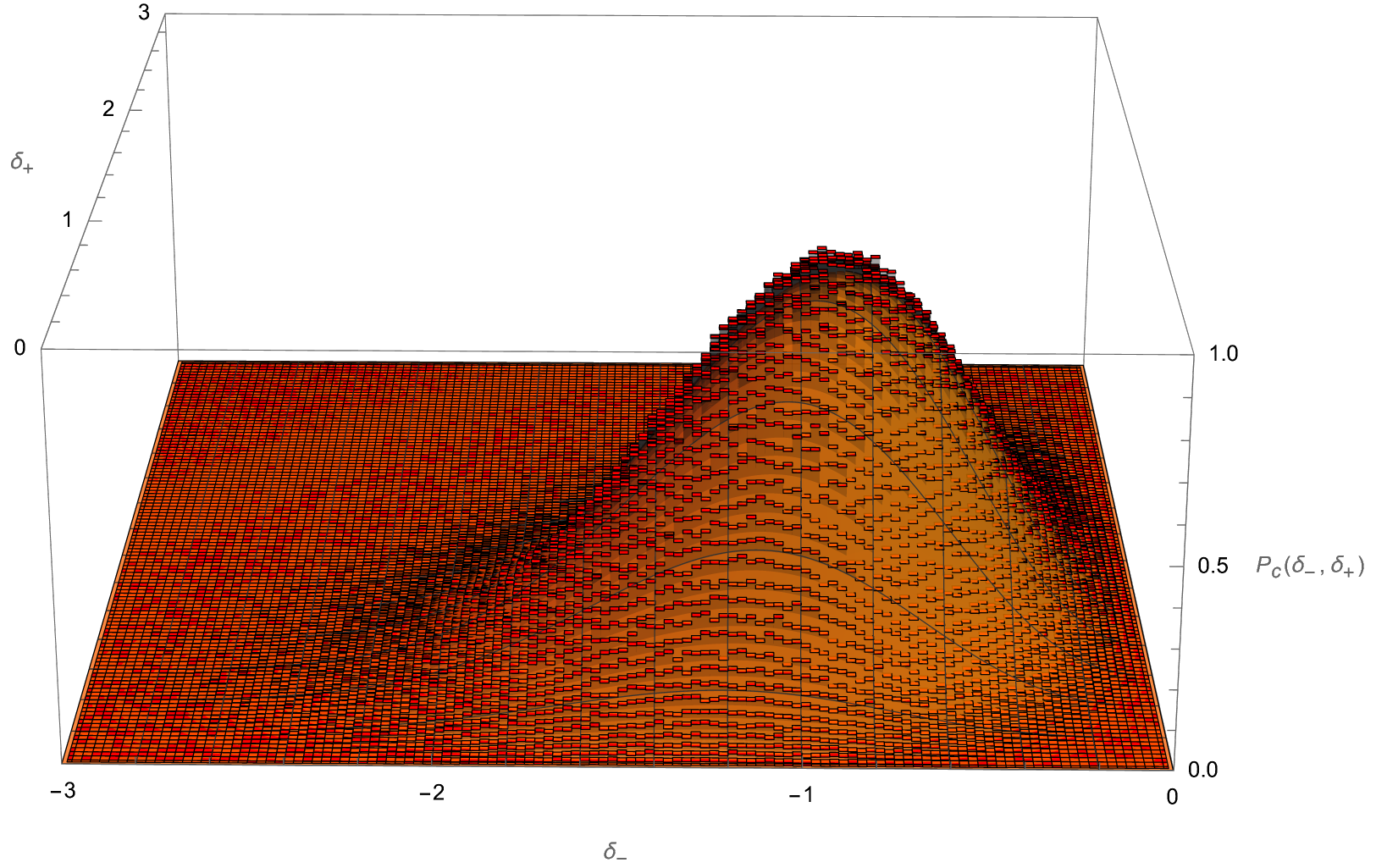}~
\includegraphics[bb=0 0 450 297,width=76mm]{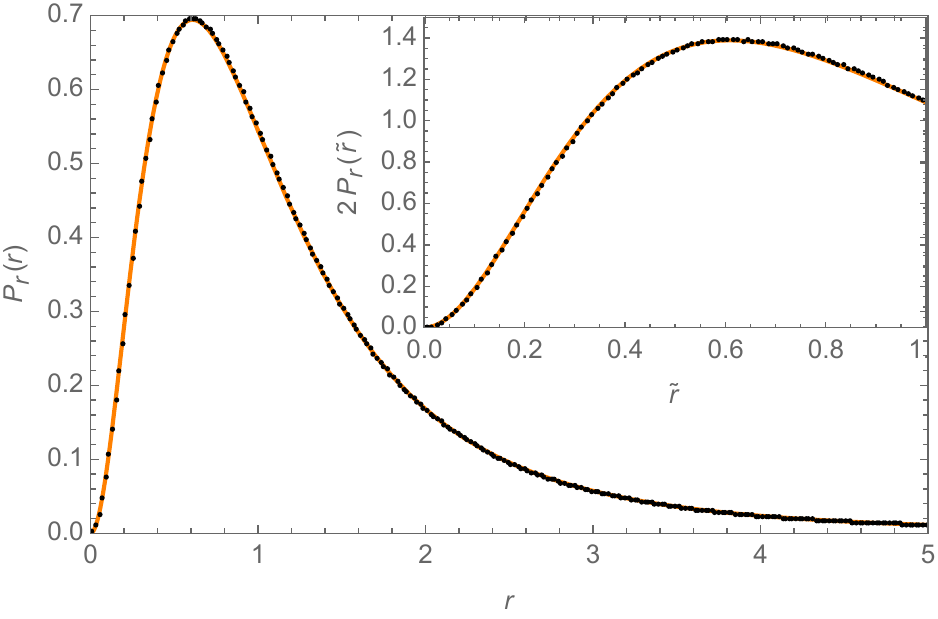}
\caption{
The joint distribution $P_{\mathrm{c}}(\delta_-,\delta_+)$ of two unfolded consecutive spacings 
$\delta_\pm=\bar{\rho}(\gamma_n)(\gamma_{n\pm 1}-\gamma_{n})$
of
$10^8$ zeroes of the Riemann $\zeta$ function $\{\frac12+i\gamma_n\}$
around $n\simeq 1.037\cdot10^{11}$ (left), 
and the distributions of the ratios 
$r_n=(\gamma_{n+1}-\gamma_{n})/(\gamma_{n}-\gamma_{n-1})$ (right)
and $\tilde{r}_n=\min(r_n, r_n^{-1})$ (right, inset).
Histograms from the Riemann zeroes (black tiles and dots) are plotted versus
the analytical results (Figs.\,1 and 2) for the GUE (orange surface and curves).
}
\end{center}
\end{figure}
Indeed, this observation has been reported in Fig.\,4 of the original article \cite{Atas:2013}
that inspired this work.
Equipped with the high precision of the moments of $\tilde{r}$
attained by our analytic derivation of $P_{\mathrm{c}}(a_1,a_2)$,
we can now revisit the rather crude observation of Ref.\,\cite{Atas:2013} 
and improve it to the level of quantifying
the systematic convergence of the distribution of the $\tilde{r}$-ratios of the Riemann zeroes
to the GUE result (\ref{Prr}).
Mean values of the moments of $\tilde{r}_n$ in four windows of zeroes 
$[\gamma_N , \gamma_{1.001 N+1}]$ for $N=10^{8},10^{9}, 10^{10}$, and  $103700788358$
are summarized in Table 1.
\begin{table}[t]
\begin{center}
\caption{Moments of $\tilde{r}_n$ for the Riemann zeroes.}
\begin{tabular}{lccccc}
\hline
$N$ & $\gamma_N$ & 
$\langle \tilde{r}_n \rangle$ & $\langle \tilde{r}^2_n \rangle$ &
$\langle \tilde{r}^3_n \rangle$ & $\langle \tilde{r}^4_n \rangle$\\
\hline
$10^8$    &  $4.265354 \cdot 10^7$ & 
0.6032357 & 0.4168926 & 0.3133507 & 0.2489623 \\
$10^9$    &  $3.718702 \cdot 10^8$ &
0.6021928 & 0.4158748 & 0.3125019 & 0.2482868\\
$10^{10}$ &  $3.293531 \cdot 10^9$ &
0.6014386 & 0.4149925 & 0.3116161 & 0.2474310\\
103700788358
& $3.058187 \cdot 10^{10}$ & 
0.6010277 & 0.4145862 & 0.3112812 & 0.2471641 \\
\hline
\end{tabular}
\end{center}
\end{table}
As the height $\gamma_N$ increases, relative deviations from Eq.\,(\ref{Er}),
$\langle \tilde{r}^k_n \rangle/ \mathbb{E}[\tilde{r}^k]-1$, 
are indeed observed to vanish systematically from above, 
heuristically in proportion to $\bar{\rho}(\gamma_N)^{-3}$,
indicating an approach to complete quantum chaoticity
in the ``thermodynamic limit" $\gamma_N\to \infty$ (Fig.\,4).
\begin{figure}[ht] 
\begin{center}
\includegraphics[bb=0 0 406 256,width=80mm]{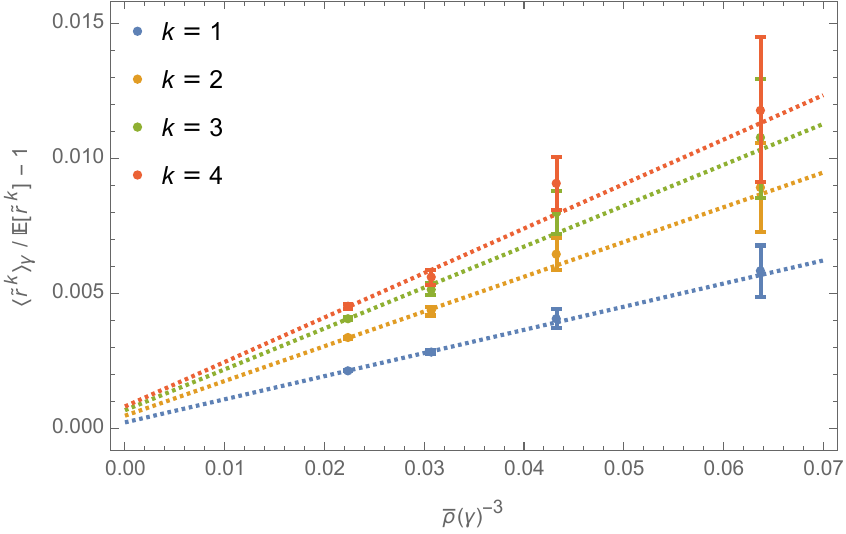}
\caption{Relative deviations of the moments of the $\tilde{r}$-ratio
$\langle\tilde{r}^k_n\rangle\ (k=1,\ldots,4)$
in four windows of the Riemann zeroes $\{\frac12+i \gamma_n\}$,
$n\in [N, 1.001N+1]$ for $N=10^{8}, 10^{9}, 10^{10}$, and $\simeq 10^{11}$,
from the GUE results $\mathbb{E}[\tilde{r}^k]$, plotted versus
$\bar{\rho}(\gamma_N)^{-3}$.
Error bars represent statistical fluctuations due to an arbitrary choice of windows,
estimated by the Jackknife method of splitting each data set in 10 bins.
Dotted lines are linear fits with the least $\chi^2$
for each set of four points in the same color.
}
\end{center}
\end{figure}

We remark that the systematic deviations of the statistics of the Riemann zeroes
from the GUE (at infinite $N$) were previously studied in
Refs.\,\cite{Bogomolny:2006,Forrester:2015,Bornemann:2017}.
There, the ``finite-size corrections" in various statistical distributions
for the Riemann zeroes were reported to agree well with those for
the circular unitary ensemble (CUE)
at finite $N_{\mathrm{eff}}\approx 1.446124 \cdot\bar{\rho}(\gamma_N)$.
The relationship between these observations and our finding above
will need to be clarified.

\section*{Supplementary materials}
{\tt Janossy\_TW\_N.nb}:
Mathematica Notebook for generating $J_{1}(0;[a_1, a_2])$
analytically by the TW system (Eqs.\,(\ref{ode}) and (\ref{bc}))
or numerically by the Nystr\"{o}m-type approximation (\ref{Nystrom}).\\
{\tt Janossy\_Sin.dat}:
numerics of $J_1(0; [a_1,a_2])$ in the range
$(a_1,a_2)\in [-20, 0]\times[0,20]$, 
provided upon request by email.

\section*{Funding}
This work is supported by Japan Society for the Promotion of Science (JSPS) Grants-in-Aids
for Scientific Research (C) No.\,7K05416.

\section*{Acknowledgments}
I thank the L-functions and modular forms database (LMFDB) Collaboration \cite{LMFDB}
for making the data of the Riemann
$\zeta$ function zeroes publicly downloadable,
and Peter Forrester for an informative comment on the manuscript.

\end{document}